# The Dynamic Behavior (Effects) of Impulsive Toxicant Input on a Single-Species Population in a Small Polluted Environment

Li Dongmei[1]   Yue Wu[2]   Xu Yajing[1]

(1 Applied Science College, Harbin University of Science and Technology, Harbin 150080 china
  2 Corresponding Author, Schlumberger WesternGeco, 3519 Heartland Key LN, Katy, TX, 77494)

**Abstract:** In this paper, we study a single-species population model with pulse toxicant input to a small polluted environment. The intrinsic rate of population is affected by environment and toxin in organisms. The toxin in organisms is influenced by toxin in environment and the food chain. A new mathematical model is derived. By the Pulse Compare Theorem, we find the surviving threshold of the population and obtain the sufficient conditions of persistence and extinction of the population.

## 1  Introduction

In the real world, with the rapid development of modern industry and agriculture, environmental pollution has become an increasingly serious problem. Untreated pollutants are release to the environment continuously. This causes many serious environmental problems and damages the ecological system.   This   is a global issue. Populations of a lot of animal species have become extinct or endangered. Therefore, it is very important to study  surviving conditions of the population in polluted environment.

In the 1980s, Hallam et al. (1983-1984) studied the toxicant effectsin the environment on a single-species population. In their paper, they assumed that, relative to population size, the capacity of the environment is large, so the population absorption and excretion of toxins can be omitted. Many good results were obtained about the extinction and persistence of the population[1-3]. But in a relatively closed environment, the effect of population's own emission toxin can't be omitted. He Jinwei (usually we write He ) et al.(2007,2009) studied the survival problem of the population assuming that the intrinsic rate of population is to be affected by the environment and the toxins of the body[4-7].

In most cases, it is assumed that it is a continuous process for the toxins entering into the environment But in the real world, it is not always true, and the majority of cases are often represented as a periodic emission, such as industrial waste water or waste water discharge, agricultural pesticide spraying, etc. In these cases, the discharge time of toxins, compared with the population's life cycle, is very short, but their effect on the organism is of longer term. Liu Bing (May Liu), et al. (2003)

studied the survival effects of population on the pulse cycle toxin emissions under a fixed input quantitative toxin [8] at a fixed time. Zhang Hong, et al. (2008) established a single population model in a polluted environment by assuming that other external toxins discharged into the environment at a fixed time period. They showed that the population is extinct, when the pulse period is less than a certain threshold; conversely, the population is permanent. They also demonstrated that sustained living conditions can ensure existence and uniqueness of positive periodic solutions which are globally and asymptotically stable [9,10]. Jiao Jianjun et al.(2009) investigated a single population model with impulsive toxicant input in a polluted environment under the assumption that the toxins of population are also affected by the toxins in the food chain, discussed the extinction and permanent existence of the population, and drawed a conclusion that the population can be protected by changing the input toxin quantity and period [11-13].

On the basis of the work in [11], we establish a model by assuming that the capacity of the environment is not large enough, and the intrinsic rate of population is affected by the environment and the toxins level in the body. We also assume that the toxicant stored by the dead organisms is considered as one type of toxins. Then we have the following model

$$\begin{cases} \dfrac{dx(t)}{dt} = x(t)\left(r_0 - \alpha C_o(t) - \beta C_e(t) - \lambda x(t)\right) \\ \dfrac{dC_o(t)}{dt} = KC_e(t) + fC_e(t) - \left(g + m + b_0 - \lambda x(t)\right)C_o(t) \\ \dfrac{dC_e(t)}{dt} = \left[-K_1 C_e(t) + \left(g_1 + d_1 + \alpha_1 C_o(t) + \beta_1 C_e(t)\right)C_o(t)\right]x(t) - hC_e(t) \end{cases} \quad t \neq nT \quad (1)$$

$$\Delta x(t) = 0, \ \Delta C_o(t) = 0, \ \Delta C_e(t) = \mu, \ t = nT$$

In model (1), $x(t)$ is the population biomass at time $t$; $C_o(t)$ denotes the concentration of the toxicant in an organism at time, $t$ and $C_e(t)$ represents the concentration of the toxicant in the environment at time $t$; $\lambda$ is intra specific competition of the population; $r_0 = b_0 - d_0$ is the intrinsic growth of the population $x(t)$ without toxicant, where $b_0$ is the growth rate of the population $x(t)$ without toxicant, $d_0$ is the death rate of the population $x(t)$ without toxicant, and $b_0 > r_0$. $t \in R^+, n \in N$, where $N$ is a non-negative integer set. Due to the erosion of toxins in the environment, intrinsic rate of population is considered to be linear dose concentration-response functions of the population's and environmental toxins $r_0 - \alpha C_o(t) - \beta C_e(t)$; $KC_e(t)$ is the uptake rate of the toxicant from the environment by the population at time $t$; $fC_e(t)$ is the uptake rate of the toxicant from the food chain by the population at time $t$; $gC_o(t)$ and $mC_o(t)$ are the egestion and depuration rates of the toxicant in the organism, respectively; $hC_e(t)$ is the toxicant loss from the environment itself by volatilization and so on; and

$$g_1 = \frac{gm_0}{m_e}, \quad K_1 = \frac{Km_0}{m_e}, \quad d_1 = \frac{d_0 m_0}{m_e}, \quad \alpha_1 = \frac{\alpha m_0}{m_e}, \quad \beta_1 = \frac{\beta m_0}{m_e}. \tag{2}$$

Where $m_0$ represents the average mass of each individual of the population, $m_e$ represents the total mass of the medium in the environment; $\mu$ represents the toxicant input amount at each cycle; $T$ represents the period of the impulsive effect for the exogenous input of toxicant ;where $g_1$, $K_1$, $r_0$, $b_0$, $d_1$, $g$, $m$, $K$, $h$ are positive constants.

In this paper, we study the dynamic behavior of model (1). In Section 2, we prove that the model (1) has the non-negative solutions and they are ultimately bounded by inequality scaling method, thus the survival upper bound of the population is found. In Section 3, by the Pulse Compare Theorem, we get the solution of model (1), which has a non-negative lower bound and derive the sufficient condition of persistent survival of the population. In Section 4, we obtain the sufficient condition for extinction of the population. Some summaries are given in the last Section.

## 2  Positive and Boundedness of Solutions

In order to prove the persistence of solutions for model (1), we need to show that they are non-negative and have upper and lower bounds. First we prove that there does exist a positive solution.

We set initial values of model (1) as followings:
$$x(0) > 0, \quad 0 \le C_o(0) \le 1, \quad 0 \le C_e(0) \le 1. \tag{3}$$

In model(1), we assume $t \in R_+ = [0, \infty)$, $n \in N$, where $N$ is a non-negative integer set.

First, we have the following conclusion with respect to the positive property of solutions of model (1).

**Theorem 1** The solution $(x(t), C_o(t), C_e(t))$ of model (1) with initial conditions (3) is non-negative.

**Proof.** Integrating the first function of model (1) from $0$ to $t$ gives
$$x(t) = x(0) \exp \int_0^t \left( r_0 - \alpha C_o(\tau) - \beta C_e(\tau) - \lambda x(\tau) \right) d\tau.$$
So, if $x(0) > 0$, we have $x(t) > 0$ for $t \ge 0$.

Next, we prove $C_o(t) \ge 0$, $C_e(t) \ge 0$.

Since $C_e(0) \ge 0$, $\Delta C_e(t) = \mu > 0$, it is obvious $C_e(0^+) > 0$. But, for $C_o(0)$, we have two cases as follows.

Case I: $C_o(0) = 0$.

As $\Delta C_o(t) = 0$, from the second and third functions of model (1), we have
$$\left. \frac{dC_o(t)}{dt} \right|_{t=0^+} = KC_e(0^+) + fC_e(0^+) - \left(g + m + b_0 - \lambda x(0^+)\right) C_o(0^+) = (K+f) C_e(0^+) > 0$$

$$\left.\frac{dC_e(t)}{dt}\right|_{t=0^+} = -K_1 C_e(0^+) x(0^+) - h C_e(0^+) < 0.$$

Hence there must exist a positive number $\varepsilon$ such that $C_o(t) > 0$, $C_e(t) > 0$ for $t \in (0, \varepsilon)$.

Then, when $t > 0$, we have
$$C_o(t) > 0, \quad C_e(t) > 0. \tag{4}$$

If it is not, there must exist a positive number $t^*$ such that $C_o(t^*) \cdot C_e(t^*) = 0$ for $t^* \in ((n-1)T, nT]$ and $C_o(t) > 0$, $C_e(t) > 0$ for $t \in (0, t^*)$. Therefore there are only three cases at the end point $t^*$:

For the first case: $C_o(t^*) = 0$, $C_e(t^*) > 0$.

If $C_o(t) > 0$ is true, then it is obvious $\dfrac{dC_o(t^*)}{dt} \leq 0$ for $t \in (0, t^*)$. But from the second function of model (1), we have
$$\left.\frac{dC_o(t)}{dt}\right|_{t=t^*} = (K+f) C_e(t^*) > 0.$$

There is a contradiction, so the first situation doesn't hold.

For the second case: $C_o(t^*) > 0$, $C_e(t^*) = 0$.

If $C_e(t) > 0$ is true, then it is obvious $\dfrac{dC_e(t^*)}{dt} \leq 0$ for $t \in (0, t^*)$. But from the third function of model (1), we have
$$\left.\frac{dC_e(t)}{dt}\right|_{t=t^*} = \left[g_1 + d_1 + \alpha_1 C_o(t^*)\right] C_o(t^*) x(t^*) > 0.$$

There is a contradiction, thus the second situation is not true.

For the third case: $C_o(t^*) = 0$, $C_e(t^*) = 0$.

It is obvious that $(x(t), 0, 0)$ is the solution of model (1). At the same time, it is also the solution of model (1) with initial values satisfying $x(t^*) > 0$, $C_o(t^*) = 0$, $C_e(t^*) = 0$. The uniqueness theorems of solution yields $C_o(t) \equiv 0$, $C_e(t) \equiv 0$ for $t > 0$. This is also a contradiction. Hence the third situation doesn't hold. We conclude that $C_o(0) = 0$.

Case II: $C_o(0) > 0$.

From $C_o(0) > 0$ and the continuity of $C_o(t)$, for any $\varepsilon_1 > 0$, we have $C_o(t) > 0$ for $t \in (0, \varepsilon_1)$.

Furthermore, $C_e(0^+) > 0$, whatever $\left.\dfrac{dC_e(t)}{dt}\right|_{t=0^+}$ is positive or negative, we can promise that, for any $\varepsilon_2 > 0$, we have $C_e(t) > 0$ for $t \in (0, \varepsilon_2)$.

So let $\varepsilon = \min(\varepsilon_1, \varepsilon_2) > 0$, there is $C_o(t) > 0$, $C_e(t) > 0$ for $t \in (0, \varepsilon)$.

Next we prove, when $t > 0$, there is

$$C_o(t) > 0, \quad C_e(t) > 0.$$

Then the proof of Case II is similar to that of Case I, the result still holds.  □

Next we prove that all positive solutions of model (1) have upper bounds.

**Theorem 2** For model (1), if $\dfrac{fr_0}{\lambda} < \dfrac{(h-m)g}{g_1}$, for any $t \in R^+$, there must exist a positive number $M$, such that

$$\limsup_{t \to \infty} x(t) \le M, \quad \limsup_{t \to \infty} C_o(t) \le M, \quad \limsup_{t \to \infty} C_e(t) \le M.$$

**Proof.** From Theorem 1 and the first equation of model (1), we have

$$\frac{dx(t)}{dt} \le x(t)(r_0 - \lambda x(t)).$$

Standard comparison Theorem produces

$$\limsup_{t \to \infty} x(t) \le \frac{r_0}{\lambda} @ M_1. \tag{5}$$

Defining $V(t) = C_o(t)x(t) + \dfrac{g}{g_1} C_e(t)$ and using (2), one can obtain

$$D^+V(t) + mV(t) = fC_e(t)x(t) - (h-m)\frac{g}{g_1}C_e(t).$$

Expression (5) and $\dfrac{fr_0}{\lambda} < \dfrac{(h-m)g}{g_1}$, when $t \ne nT$, gives

$$D^+V(t) + mV(t) \le \left(\frac{fr_0}{\lambda} - (h-m)\frac{g}{g_1}\right)C_e(t) < 0.$$

When $t = nT$, there is $V(nT^+) = V(nT) + \dfrac{\mu g}{g_1}$, so for $t \in (nT, (n+1)T]$, pulse inequality (Lemma 2.2) in [14] gives

$$V(t) \le V(0)e^{-mt} + \frac{\mu g}{g_1}\frac{e^{-m(t-T)}}{1 - e^{-mT}} + \frac{\mu g}{g_1}\frac{e^{mT}}{e^{mT} - 1}.$$

Hence $V(t)$ is uniformly bounded, which is

$$\limsup_{t \to \infty} V(t) \le \frac{\mu g}{g_1}\frac{e^{mT}}{e^{mT} - 1} @ M_2.$$

According to the definition of $V(t)$ and theorem 1, one can derive

$$\limsup_{t \to \infty} C_o(t)x(t) \le M_2, \quad \limsup_{t \to \infty} C_e(t) \le \frac{g_1}{g}M_2. \tag{6}$$

The second function (or equation) of model (1) and (6) leads to

$$\limsup_{t \to \infty} C_e(t) \le \frac{(K+f)g_1 M_2 + \lambda M_2 g}{g(g + m + b_0)} @ M_3.$$

Let $M = \max\left(M_1, \dfrac{g_1}{g}M_2, M_3\right)$, so there is

$$\limsup_{t \to \infty} x(t) \le M, \quad \limsup_{t \to \infty} C_o(t) \le M, \quad \limsup_{t \to \infty} C_e(t) \le M.$$

From Theorem 1 and Theorem 2, there is a invariant set in mode (1); that is

$$\Omega = \{(x(t), C_o(t), C_e(t)) | 0 \leq x(t) \leq M, 0 \leq C_o(t) \leq M, 0 \leq C_e(t) \leq M\}.$$

## 3 Persistence Survival of Population

In Theorem 2, we know that the solutions of model (1) have upper bounds. In this section, in order to investigate the survival of the population, we will prove that model (1) has a non-negative lower bound. Now we can analyze the model (1) by the Impulsive Differential Equations Comparison Theorem to find the lower bound as follows.

**Theorem 3** For model (1), if

$$r_0 T > \frac{\alpha \mu (K+f)}{h(g+m+b_0)} + \frac{\beta \mu}{h}, \tag{7}$$

then the population $x(t)$ will be uniformly persistent.

**Proof** From Theorem 2, we know that $x(t)$ is ultimately bounded. Hence, in order to prove that the population $x(t)$ is uniformly persistent, we can only need to show that $x(t)$ has the lower bound. If not, for any $\delta > 0$, when $t > 0$, there is

$$x(t) < \delta. \tag{8}$$

Considering the last two equations of model (1):

$$\begin{cases} \dfrac{dC_o(t)}{dt} = (K+f)C_e(t) - (g+m+b_0)C_o(t) + \lambda x(t)C_o(t) \\ \dfrac{dC_e(t)}{dt} = -hC_e(t) + \left[-K_1 C_e(t) + (g_1 + d_1 + \alpha_1 C_o(t) + \beta_1 C_e(t))C_o(t)\right]x(t) \end{cases} t \neq nT$$

$$\Delta C_o(t) = 0, \ \Delta C_e(t) = \mu, \ t = nT \tag{9}$$

We set up $(C_o(t), C_e(t))$ is the solution of model (9).

From Theorem 2 and (8), we get the pulse comparison equation corresponding equation of model (9):

$$\begin{cases} \dfrac{du(t)}{dt} = (K+f)v(t) - (g+m+b_0)u(t) + \lambda M \delta \\ \dfrac{dv(t)}{dt} = -hv(t) + a \end{cases} t \neq nT,$$

$$\Delta u = 0, \ \Delta v = \mu, \ t = nT \tag{10}$$

where $a = (g_1 + d_1 + \alpha_1 M + \beta_1 M) M \delta$.

Let $(u(t), v(t))$ be the solution of model (10) with initial values $u(0) = C_o(0)$, $v(0) = C_e(0)$.

We will finish the proof by the following three steps.

-First we prove that model (10) only has a positive periodic solution $\left(\bar{u}(t), \bar{v}(t)\right)$, which is globally attractive.

In the interval $(nT, (n+1)T]$, the solution of the second function of model (10) is

$$v(t) = \frac{a}{h} + \left(v(nT^+) - \frac{a}{h}\right)e^{-h(t-nT)}. \qquad (11)$$

From $\Delta v(t) = \mu$ and (11), we have

$$v((n+1)T^+) = \left(v(nT^+) - \frac{a}{h}\right)e^{-hT} + \frac{a}{h} + \mu, \qquad (12)$$

which implies a stroboscopic map

$$v((n+1)T^+) = H\left(v(nT^+)\right) \qquad (13)$$

where

$$H(y) = \left(y - \frac{a}{h}\right)e^{-hT} + \frac{a}{h} + \mu.$$

We get the only fixed point of map (13); that is

$$v^* = \frac{\mu}{1 - e^{-hT}} + \frac{a}{h}. \qquad (14)$$

It is easily to show that $|H'(v^*)| = e^{-hT} < 1$, so sequence $\{v((n+1)T^+)\}$ converges to $v^*$.

Using (11) and (14), we have

$$\bar{v}(t) = \frac{a}{h} + \left(v^* - \frac{a}{h}\right)e^{-h(t-nT)}, \quad nT < t \leq (n+1)T$$

Similarly, from the first function of model (10), we can get

$$\bar{u}(t) = p + \frac{q(v^* - a/h)}{\mu}e^{-h(t-nT)} + \left(u^* - p - \frac{q(v^* - a/h)}{\mu}\right)e^{-(g+m+b_0)(t-nT)}$$

where, $p = \dfrac{(K+f)a/h + \lambda M \delta}{g + m + b_0}$, $q = \dfrac{(K+f)\mu}{g + m + b_0 - h}$

$$u^* = p + \frac{q(v^* - a/h)}{\mu}\left(\frac{e^{-hT} - e^{-(g+m+b_0)T}}{1 - e^{-(g+m+b_0)T}}\right).$$

Therefore the model (10) only has a positive periodic solution $(\bar{u}(t), \bar{v}(t))$.

Now we prove that the positive periodic solution $(\bar{u}(t), \bar{v}(t))$ is also globally asymptotically stable.

If $(u(t), v(t))$ is the any solution of model (10), define the transformation

$$M(t) = u(t) - \bar{u}(t), \quad N(t) = v(t) - \bar{v}(t), \qquad (15)$$

Then expression (10) is changed as follows:

$$\begin{cases} \dfrac{dM(t)}{dt} = (K+f)N(t) - (g+m+b_0)M(t) \\ \dfrac{dN(t)}{dt} = -hN(t) \end{cases} \qquad (16)$$

Let $M(0) = u(0)$, $N(0) = v(0)$ be initial values of model (16).

From the second function of model (16), we have
$$N(t) = v(0)e^{-ht} \tag{17}$$

First function of model (16) also gives
$$M(t) = \frac{(K+f)v(0)}{h-(g+m+b_0)}e^{-ht} + \left(u(0) - \frac{(K+f)v(0)}{h-(g+m+b_0)}\right)e^{-(g+m+b_0)t} \tag{18}$$

Using (17) and (18), we get $\lim_{t \to \infty} N(t) = 0$, $\lim_{t \to \infty} M(t) = 0$. So $(\bar{u}(t), \bar{v}(t))$ is globally attractive.

-Second we prove the population $x(t)$ is uniformly persistent.

Using the Comparison Theorem[15] and the globally asymptotically stable property of $(\bar{u}(t), \bar{v}(t))$, there exists a positive number $T_0 > 0$, for arbitrarily small $\varepsilon > 0$, and when $t > T_0$, we have
$$\begin{cases} C_o(t) \leq u(t) \leq \bar{u}(t) + \varepsilon \\ C_e(t) \leq v(t) \leq \bar{v}(t) + \varepsilon \end{cases} \tag{19}$$

Using (19) and the first function of model (1), we have
$$\frac{dx(t)}{dt} = x(t)(r_0 - \alpha C_o(t) - \beta C_e(t) - \lambda x(t)) \tag{20}$$
$$\geq x(t)(r_0 - \alpha(\bar{u}(t) + \varepsilon) - \beta(\bar{v}(t) + \varepsilon) - \lambda \delta)$$

Define $n_1 \in N$ and $n_1 T > T_0$. Integrating (20) from $nT$ to $(n+1)T$ ($n \geq n_1$) leads to
$$x((n+1)T) \geq x(nT)\exp\int_{nT}^{(n+1)T}(r_0 - \alpha(\bar{u}(t) + \varepsilon) - \beta(\bar{v}(t) + \varepsilon) - \lambda \delta)dt$$
$$= x(nT)\exp\left[(r_0 - \lambda\delta - \alpha\varepsilon - \beta\varepsilon)T - \alpha\left(pT - \frac{q}{h(1-e^{-hT})}(e^{-hT} - 1)\right.\right.$$
$$\left.\left. - \frac{1}{g+m+b_0}\left(u^* - p - \frac{q}{1-e^{-hT}}\right)(e^{-(g+m+b_0)T} - 1)\right) - \beta\left(\frac{aT}{h} - \frac{v^* - a/h}{h}(e^{-hT} - 1)\right)\right]$$
$$= x(nT)\exp\left[(r_0 - \lambda\delta - \alpha\varepsilon - \beta\varepsilon)T - \alpha\left(pT + \frac{q}{h} - \frac{q}{g+m+b_0}\right) - \beta\left(\frac{aT}{h} + \frac{\mu}{h}\right)\right]$$
$$= x(nT)\exp(\kappa) \geq x(0^+)\exp(n\kappa) \tag{21}$$

where
$$\kappa = (r_0 - \lambda\delta - \alpha\varepsilon - \beta\varepsilon)T - \alpha\left(pT + \frac{q}{h} - \frac{q}{g+m+b_0}\right) - \beta\left(\frac{aT}{h} + \frac{\mu}{h}\right)$$
$$= \left(r_0 - \lambda\delta - \alpha\varepsilon - \beta\varepsilon - \alpha p - \frac{a\beta}{h}\right)T - \alpha\left(\frac{q}{h} - \frac{q}{g+m+b_0}\right) - \beta\frac{\mu}{h}$$
$$= \left(r_0 - \lambda\delta - \alpha\varepsilon - \beta\varepsilon - \frac{\alpha(K+f)a/h + \lambda\alpha M\delta}{g+m+b_0} - \frac{a\beta}{h}\right)T$$
$$- \frac{\alpha^2\mu(K+f)}{h(g+m+b_0)} - \frac{\beta\mu}{h} \tag{22}$$

From (7), we know $r_0 T > \dfrac{\alpha\mu(K+f)}{h(g+m+b_0)} + \dfrac{\beta\mu}{h}$. Hence, for given $M > 0$, we choose sufficiently small $\delta > 0$ and $\varepsilon > 0$, such that

$$\left(r_0 - \alpha\varepsilon - \beta\varepsilon - \lambda\delta - \dfrac{\alpha(K+f)a/h + \lambda\alpha M\delta}{g+m+b_0} - \dfrac{a\beta}{h}\right)T > \dfrac{\alpha\mu(K+f)}{h(g+m+b_0)} + \dfrac{\beta\mu}{h}.$$

From (22), we get $\kappa > 0$. So (21) yields $\lim_{n\to\infty} x(nT) = \infty$. This is a contradiction with (8). Hence there exists a positive number $t_1 \geq T_0$ such that $x(t_1) > \delta$.

-Third we prove, when $t \geq t_1$, we have

$$x(t) \geq \delta \exp(-\omega T), \tag{23}$$

where $\omega = \sup_{nT < t \leq (n+1)T} \{|r_0 - \alpha(\bar{u}(t)+\varepsilon) - \beta(\bar{v}(t)+\varepsilon) - \lambda\delta|\}$.

If not, there exist $t_2 > t_1$ such that $x(t_2) < \delta\exp(-\omega T)$. According to the continuity of $x(t)$ with $t$, there exists $t^* \in (t_1, t_2)$ such that $x(t^*) = \delta$ and $x(t) < \delta$ for $t \in (t^*, t_2)$. For any $t \in (t^*, t_2)$, we choose a non-negative integer $l$ such that $t_2 \in (t^* + lT, t^* + (l+1)T]$. Integrating (20) from $t^*$ to $t_2$, we get

$$\delta\exp(-\omega T) > x(t_2) \geq x(t^*)\exp\left(\int_{t^*}^{t_2} (r_0 - \alpha(\bar{u}(t)+\varepsilon) - \beta(\bar{v}(t)+\varepsilon) - \lambda\delta)dt\right)$$

$$= \delta\exp\left(\left(\int_{t^*}^{t^*+lT} + \int_{t^*+lT}^{t_2}\right)(r_0 - \alpha(\bar{u}(t)+\varepsilon) - \beta(\bar{v}(t)+\varepsilon) - \lambda\delta)dt\right)$$

$$> \delta\exp\left(\int_{t^*+lT}^{t_2} (r_0 - \alpha(\bar{u}(t)+\varepsilon) - \beta(\bar{v}(t)+\varepsilon) - \lambda\delta)dt\right)$$

$$\geq \delta\exp(-\omega T)$$

There is a contradiction. So (23) is right.

In summary, we have

$$\liminf_{t\to\infty} x(t) \geq \delta\exp(-\omega T).$$

□

Sufficient conditions of persistence of the population are obtained from Theorem 3. Conversely, if the condition of Theorem 3 is false, the population may be extinct. In next section we study the conditions of the extinction of the population.

## 4 Extinction of Population

**Theorem 4** For model (1), if

$$r_0 T \leq \dfrac{\alpha\mu(K+f)}{h(g+m+b_0)} + \dfrac{\beta\mu}{h}, \tag{24}$$

Then we have the population $x(t)$ becomes extinct.

**Proof.** For proving the extinction of population $x(t)$, we only prove $\lim_{t\to\infty} x(t) = 0$.

Using Theorem 2 and the last two functions of model (1), we have

$$\begin{cases} \dfrac{dC_o(t)}{dt} \geq KC_e(t) + fC_e(t) - (g+m+b_0)C_o(t) \\ \dfrac{dC_e(t)}{dt} \geq -(K_1M+h)C_e(t) \end{cases} t \neq nT \qquad (25)$$
$$\Delta C_0(t) = 0, \ \Delta C_e(t) = \mu, \ t = nT$$

The pulse comparison equation corresponding to the model (25) is

$$\begin{cases} \dfrac{ds(t)}{dt} = Kw(t) + fw(t) - (g+m+b_0)s(t) \\ \dfrac{dw(t)}{dt} = -(K_1M+h)w(t) \end{cases} t \neq nT \qquad (26)$$
$$\Delta s(t) = 0, \ \Delta w(t) = \mu, \ t = nT$$

Let $(s(t), w(t))$ denote the solution of model (26) with the initial values $s(0) = C_o(0)$, $w(0) = C_e(0)$.

Following the same procedure as the solving process of model (10), model (26) has a positive periodic solution as follows:

$$\begin{cases} \overline{s}(t) = s^* e^{-(g+m+b_0)(t-nT)} + \dfrac{\mu(K+f)(e^{-(g+m+b_0)(t-nT)} - e^{-(K_2M+h)(t-nT)})}{(K_2M+h-g-m-b_0)(1-e^{-(K_2M+h)T})} \\ \overline{w}(t) = w^* e^{-(K_2M+h)(t-nT)} \end{cases}$$

where

$$\begin{cases} s^* = \dfrac{\mu(K+f)(e^{-(g+m+b_0)T} - e^{-(K_2M+h)T})}{(K_2M+h-g-m-b_0)(1-e^{-(K_2M+h)T})(1-e^{-(g+m+b_0)T})} \\ w^* = \dfrac{\mu}{1-e^{-(K_2M+h)T}} \end{cases}$$

And the positive periodic solution $(\overline{s}(t), \overline{w}(t))$ of (26) is globally asymptotically stable.

Using the Comparison Theorem [15] and the globally asymptotically stable property of $(\overline{s}(t), \overline{w}(t))$, there exists a positive number $t_0 > 0$, for arbitrarily small $\varepsilon_0 > 0$, and when $t > t_0$, we have

$$C_o(t) \geq s(t) \geq \overline{s}(t) - \varepsilon_0, \ C_e(t) \geq w(t) \geq \overline{w}(t) - \varepsilon_0. \qquad (27)$$

For proving the extinction of the population $x(t)$ of model (1), the contradiction method is used. Assuming that, for arbitrarily small $\eta > 0$, when $t > t_0$, there is

$$x(t) \geq \eta. \qquad (28)$$

Using (27), (28) and the first function of model (1), we get, when $t > t_0$

$$\dfrac{dx(t)}{dt} \leq x(t)\left(r_0 - \alpha\left(\overline{s}(t) - \varepsilon_0\right) - \beta\left(\overline{w}(t) - \varepsilon_0\right) - \lambda\eta\right) \qquad (29)$$

Setting $n_2 \in N$ and $n_2T > t_0$, and integrating (29) from $nT$ to $(n+1)T$ $(n \geq n_2)$ yields to

$$x((n+1)T) \leq x(nT)\exp\left(\int_{nT}^{(n+1)T}(r_0 - \alpha(\bar{s}(t)-\varepsilon_0) - \beta(\bar{w}(t)-\varepsilon_0) - \lambda\eta)dt\right)$$

$$= x(nT)\exp\left((r_0 + \alpha\varepsilon_0 + \beta\varepsilon_0 - \lambda\eta)T - \frac{\alpha\mu(K+f)}{(g+m+b_0)(K_2M+h)} - \frac{\beta\mu}{K_2M+h}\right)$$

$$= x(nT)\exp(\kappa_1) \leq x(0^+)\exp(n\kappa_1) \tag{30}$$

where

$$\kappa_1 = (r_0 + \alpha\varepsilon_0 + \beta\varepsilon_0 - \lambda\eta)T - \frac{\alpha\mu(K+f)}{(g+m+b_0)(K_1M+h)} - \frac{\beta\mu}{K_1M+h} \tag{31}$$

From (24), we know $r_0T \leq \frac{\alpha\mu(K+f)}{h(g+m+b_0)} + \frac{\beta\mu}{h}$. For given $M>0$, we choose sufficiently small $\varepsilon_0 > 0$ and $\eta > 0$ such that

$$(r_0 + \alpha\varepsilon_0 + \beta\varepsilon_0 - \lambda\eta)T \leq \frac{\alpha\mu(K+f)}{(K_1M+h)(g+m+b_0)} + \frac{\beta\mu}{K_1M+h}.$$

From (31), we get $\kappa_1 \leq 0$.

When $\kappa_1 = 0$, (30) gives $x((n+1)T) \leq 0$. From Theorem 1, we also get $x((n+1)T) \geq 0$, which means $x((n+1)T) = 0$. It demonstrates that the population $x(t)$ is eventually extinct.

When $\kappa_1 < 0$, (30) shows $x((n+1)T) \leq x(0^+)\exp(n\kappa_1) \to 0$ ($n \to \infty$), which is contradiction with (28). So there exists $t_1 > t_0$ such that $x(t_1) < \eta$.

Now we prove that, when $t > t_1$, we have

$$x(t) < \eta\exp(-\omega_1 T), \tag{32}$$

where $\omega_1 = \sup_{nT < t \leq (n+1)T}\{|r_0 - \alpha(\bar{s}(t)-\varepsilon_0) - \beta(\bar{w}(t)-\varepsilon_0) - \lambda\eta|\}$.

If not, there exists $t_2 > t_1$ such that $x(t_2) \geq \eta\exp(-\omega_1 T)$. Hence there exists $t^* \in (t_1, t_2)$ such that $x(t^*) = \eta$ and $x(t) > \eta$ for $t \in (t^*, t_2)$. We choose a non-negative integer $l_0$ such that $t_2 \in (t^* + l_0T, t^* + (l_0+1)T]$. Integrating (29) from $t^*$ to $t_2$ leads to

$$\eta\exp(\omega_1 T) \leq x(t_2) < x(t^*)\exp\left(\int_{t^*}^{t_2}(r_0 - \alpha(\bar{s}(t)-\varepsilon_0) - \beta(\bar{w}(t)-\varepsilon_0) - \lambda\eta)dt\right)$$

$$= \eta\exp\left(\left(\int_{t^*}^{t^*+l_0T} + \int_{t^*+l_0T}^{t_2}\right)(r_0 - \alpha(\bar{s}(t)-\varepsilon_0) - \beta(\bar{w}(t)-\varepsilon_0) - \lambda\eta)dt\right) \text{ There is}$$

$$\leq \eta\exp(\omega_1 T)$$

a contradiction. Therefore (32) is right. As the arbitrariness of $\eta$, we have $\lim_{t\to\infty} x(t) = 0$. □

Comments: Some connection to the next section.

## 5 Numerical Simulation

Pollutant regularly input towards the environment is directly related to the survival of the population $x(t)$. Theorem 3 and Theorem 4 give the sufficient conditions for survival and extinction of population $x(t)$. Using numerical simulation, we analysis the influence of $T$ and $\mu$ on the survival of population $x(t)$. Assuming $r_0 = 0.6$, $\alpha = 0.1$, $\beta = 0.1$, $\lambda = 0.2$, $K = 0.2$, $f = 0.1$, $g = m = 0.1$, $b_0 = 0.8$, $K_1 = 0.1$, $g_1 = 0.05$, $d_1 = 0.1$, $\alpha_1 = \beta_1 = 0.05$, $h = 0.1$, $x(0) = 1$, $C_o(0) = 0.5$, $C_e(0) = 0.8$。

Let $\mu = 1$, $T = 3$, we can get $\dfrac{\alpha\mu(K+f)}{r_0 h(g+m+b_0)} + \dfrac{\beta\mu}{r_0 h} = 2.2$, then the conditions of Theorem 3 are satisfied, population $x(t)$ is survived. As shown in Fig.1.

Let $\mu = 2$, $T = 3$, we can get $\dfrac{\alpha\mu(K+f)}{r_0 h(g+m+b_0)} + \dfrac{\beta\mu}{r_0 h} \approx 4.33$, then the conditions of Theorem 4 are satisfied, population $x(t)$ is extinct. As shown in Fig.2.

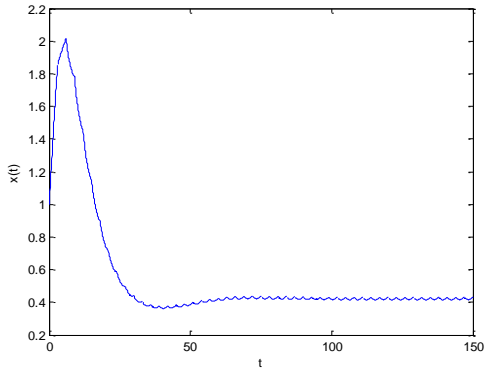 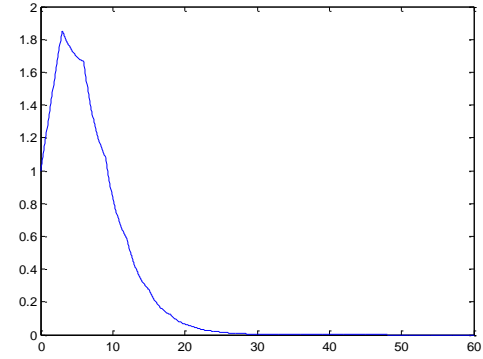

Fig.1 The extinction of population $x(t)$ when $\mu = 1$, $T = 3$.

Fig. 2 The existence of population $x(t)$ when $\mu = 2$, $T = 3$.

From Fig.1 and Fig.2, we can observe that when $T$ is the same and $\mu$ increases from 1 to 2, population $x(t)$ changes from survival to extinction.

Let $\mu = 1$, $T = 1$, we can get $\dfrac{\alpha\mu(K+f)}{r_0 h(g+m+b_0)} + \dfrac{\beta\mu}{r_0 h} = 2.2$, then the conditions of Theorem 4 are satisfied, population $x(t)$ is extinct. As shown in Fig.3.

Let $\mu = 1$, $T = 2.5$, we can get $\dfrac{\alpha\mu(K+f)}{r_0 h(g+m+b_0)} + \dfrac{\beta\mu}{r_0 h} = 2.2$, then the conditions of Theorem 3 are satisfied, population $x(t)$ is survived. As shown in Fig.4.

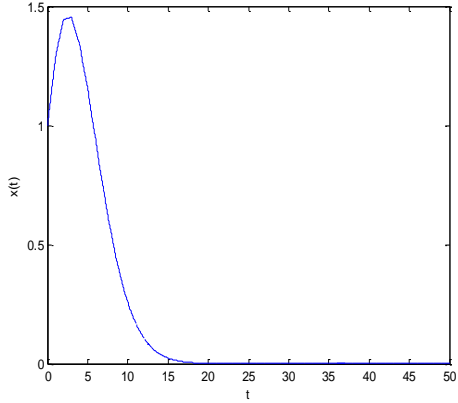
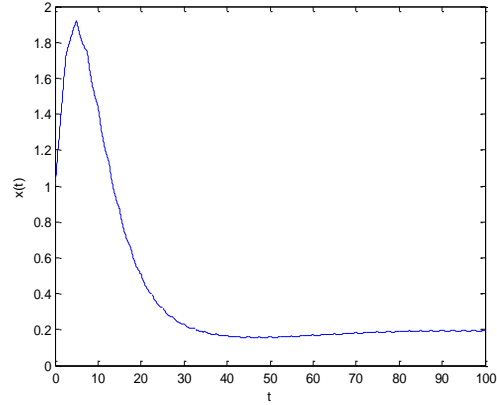

Fig. 3　The extinction of population $x(t)$ when $\mu=1$, $T=1$.

Fig. 4　The existence of population $x(t)$ when $\mu=1$, $T=2.5$.

From Fig.3 and Fig.4, we can observe that when $\mu$ is the same and $T$ increases from 1 to 2.5, population $x(t)$ changes from extinction to survival.

## 6 Conclusion

In this paper, we study a single-population model with pulse input of environmental toxin in a relatively small polluted environment. We obtain the conditions and a threshold of extinction and persistence of the population. The threshold is $R = r_0 T \Big/ \left( \dfrac{\alpha\mu(K+f)}{h(g+m+b_0)} + \dfrac{\beta\mu}{h} \right)$, that is, when $R>1$, the population is persistent; when $R \leq 1$, the population is extinct.

The degree of pollution of the environment is directly related to survival and extinction of the population. From the definition of threshold, it is learned that if the toxicant input amount is constant, we must extend the period of the exogenous input of toxicant in order to ensure the survival of the population; if the period of the exogenous input of toxicant discharge is fixed, we must decrease the toxicant input amount in view of ensuring the survival of population. At the same time, the results of numerical simulation demonstrate the influence of the period and amount of the exogenous input of toxicant on survival and extinction of populations.

Due to the limitations of the population to survive, the pollution problem in a small environment in this paper is more consistent with real problem than that in a big environment. Comparing the results of two types of environment, we can note that when the toxicant input amount is the same, the threshold of extinction and persistence of the population in a small environment is larger and the survival condition of population cecomes weaker. So in order to make the population to survive in a small environment, we only need to reduce the amount of discharge toxins and extend the time of emission. In real world, when facing pollutants from the environment, young population and adult population have different reactions. Considering the population with the different age structure has more practical significance, so this issue can be studied as a follow-up research on the basis of the current research work.

## Acknowledgments

This work was supported in part by the National Natural Science Foundation of Heilonhjiang Province under Grant no. A2016004 and The Foundation of Educational Commission of Heilongjiang Province under Grant no. A12521099.